\documentclass[%
 reprint,
 amsmath,amssymb,
 aps,
]{revtex4-2}

\usepackage{graphicx}
\usepackage{dcolumn}
\usepackage{booktabs}
\usepackage{bm}
\usepackage{placeins} 
\usepackage[version=4]{mhchem} 
\usepackage{siunitx} 
\DeclareUnicodeCharacter{03B1}{\ensuremath{\alpha}}

\begin{document}

\preprint{APS/123-QED}

\title{First-principles study of structural, electronic and magnetic properties at the \ce{(0001)Cr2O3-(111)Pt} interface}
\author{Marlies Reher}
\affiliation{Materials Theory, ETH Z\"{u}rich, Wolfgang-Pauli-Strasse 27, 8093 Z\"{u}rich, Switzerland}
\author{Nicola A. Spaldin}
\affiliation{Materials Theory, ETH Z\"{u}rich, Wolfgang-Pauli-Strasse 27, 8093 Z\"{u}rich, Switzerland}
\author{Sophie F. Weber}
\affiliation{Materials Theory, ETH Z\"{u}rich, Wolfgang-Pauli-Strasse 27, 8093 Z\"{u}rich, Switzerland}





\date{\today}

\begin{abstract}
We perform first-principles density functional calculations to elucidate structural, electronic and magnetic properties at the interface of \ce{(0001)Cr2O3-(111)Pt} bilayers. This investigation is motivated by the fact that, despite the promise of \ce{Cr2O3-Pt} heterostructures in a variety of antiferromagnetic spintronic applications, many key structural, electronic, and magnetic properties at the \ce{Cr2O3-Pt} interface are poorly understood. We first analyze all inequivalent lateral interface alignments to determine the lowest energy interfacial structure. For all lateral alignments including the lowest-energy one, we observe an accumulation of electrons at the interface between \ce{Cr2O3} and Pt. We find an unexpected reversal of the magnetic moments of the interface Cr ions in the presence of Pt compared to surface Cr moments in vacuum-terminated \ce{(0001)Cr2O3}. We also find that the heterostructure exhibits a magnetic proximity effect in the first three Pt layers at the interface with \ce{Cr2O3}, providing a mechanism by which the anomalous Hall effect can occur in \ce{(0001)Cr2O3-(111)Pt} bilayers. Our results provide the basis for a more nuanced interpretation of magnetotransport experiments on \ce{(0001)Cr2O3-(111)Pt} bilayers and should  inform future development of improved antiferromagnetic spintronic devices based on the \ce{Cr2O3-Pt} material system. 

\end{abstract}

\maketitle


\section{Introduction}
Antiferromagnetic spintronic devices offer several advantages over their ferromagnetic counterparts. Due to their zero net magnetization, antiferromagnets are robust against external magnetic field perturbations, rendering them stable against data erasure \cite{baltz_antiferromagnetic_2018, gomonay_spintronics_2014, wadley_electrical_2016}. Furthermore, the absence of a magnetic stray field for monodomain antiferromagnets allows for crosstalk-free miniaturization of bits and consequently a higher storage density \cite{baltz_antiferromagnetic_2018, han_coherent_2023, meer_antiferromagnetic_2023}.
Antiferromagnets exhibit ultrafast switching of the N\'{e}el vector, with rates in the range of THz as opposed to the GHz limit of ferromagnets \cite{fiebig_ultrafast_2008}. Finally, antiferromagnets are more ubiquitous than ferromagnets. Therefore, the exploitation of antiferromagnets in spintronic devices vastly expands the suite of candidate materials  \cite{nemec_antiferromagnetic_2018}.\\
\indent Despite their appealing properties, a major roadblock to the widespread use of antiferromagnets in practical applications is the inherent challenge of N\'{e}el vector switching and readout arising from the vanishing bulk magnetization \cite{baltz_antiferromagnetic_2018, zhang_detection_2022}. Metallic antiferromagnets with broken time-reversal, for example \ce{CuMnAs} and \ce{Mn_2Au}, can be switched via electrical current pulses \cite{wadley_electrical_2016, bodnar_writing_2018}. The underlying microscopic mechanism in this case is proposed to be a relativistic spin-orbit torque. On the other hand, insulating antiferromagnets with broken time-reversal symmetry can exhibit a magnetoelectric effect (that is, an induced bulk magnetization in response to an applied electric field, or conversely an induced bulk electric polarization in response to an applied magnetic field). The N\'{e}el vector in magnetoelectric antiferromagnets can be switched via a combination of magnetic fields and electric fields~\cite{he_robust_2010} (instead of a current) leading to lower energy consumption. In addition, it has been shown that a rigorous symmetry correspondence exists between surface magnetization (a symmetry-allowed dipole moment per unit area) and bulk magnetoelectricity, such that antiferromagnets that have a finite magnetoelectric response for an electric field along some direction $\mathbf{r}$ also have finite surface magnetization on a surface whose normal is parallel to $\mathbf{r}$ \cite{belashchenko_equilibrium_2010, weber_characterizing_2023}. Since the sign of this surface magnetization couples to the sign of the bulk N\'{e}el vector, the finite surface magnetization provides a direct method of readout \cite{he_robust_2010,wu_imaging_2011,belashchenko_equilibrium_2010,weber_characterizing_2023}.\\ 
\indent \ce{Cr2O3} in particular is an extensively studied magnetoelectric antiferromagnet with an unusually high magnetic ordering temperature of $\approx300$ $\mathrm{K}$, making it ideal for device applications \cite{dzyaloshinskii_magneto-electrical_1959, shiratsuchi_magnetoelectric_2021, shiratsuchi_magnetoelectric_2018, shiratsuchi_observation_2018, shiratsuchi_realization_2020, fallarino_magnetic_2015}. 
The roughness-robust, uncompensated surface magnetization of the \ce{(0001)Cr2O3} surface, which has the same symmetry origin as the linear magnetoelectric response for an electric field along the \ce[0001] direction, has been discussed theoretically \cite{belashchenko_equilibrium_2010, weber_characterizing_2023, weber_surface_2023} as well as experimentally \cite{cao_spin_2014, cao_surface_2015, erickson_nanoscale_2023, wu_imaging_2011}. 
An adjacent layer of Pt can in fact be used to read out the sign of surface magnetization on \ce{(0001)Cr2O3} via measuring the transverse Hall voltage of Pt in response to a current applied parallel to the interface \cite{kosub_all-electric_2015}.\\ 
\indent While prototype antiferromagnetic spintronic devices based on \ce{Cr2O3-Pt} heterostructures have been built \cite{kosub_purely_2017}, the electronic and magnetic configuration at the interface of \ce{Cr2O3} and Pt, as well as the detailed origin of the Hall voltage in \ce{Cr2O3-Pt} heterostructures are not well understood.\\ 
\indent One effect which almost certainly contributes to the Hall voltage and its dependence on the sign of surface magnetization is the so called ``Spin Hall magnetoresistance" (SMR), in which an applied in-plane electric current first generates a spin current in \ce{Pt} perpendicular to the interface via the Spin Hall effect. Then, the transverse voltage in \ce{Pt} generated due to the Inverse Spin Hall effect will have opposite sign depending on the sign of the \ce{Cr2O3} surface magnetization (and correspondingly, the sign of the N\'{e}el vector), which the spin current interacts with \cite{chen_theory_2013,schlitz_evolution_2018}.\\
\indent Another possible contributing factor is the anomalous Hall affect, or AHE. The AHE, in which an electric current develops transverse to an applied current, is symmetry-forbidden in both bulk \ce{Cr2O3} and nonmagnetic bulk fcc \ce{Pt}. Note that \ce{Cr2O3} in fact has broken time-reversal symmetry in the bulk, which is the minimal symmetry breaking required for the AHE, but in this case the AHE must still vanish due the product of inversion and time-reversal symmetry being preserved, although each symmetry individually is broken. However, the AHE has been reported to be present in \ce{Cr2O3-Pt} heterostructures \cite{kosub_all-electric_2015, wang_magnetic-field_2022, cheng_evidence_2019, moriyama_giant_2020}. A finite Pt magnetization induced in a \ce{(0001)Cr2O3-(111)Pt} heterostructure via a magnetic proximity effect \cite{cao_magnetization_2017, kosub_all-electric_2015} would break time-reversal symmetry in Pt, thus allowing for a nonvanishing AHE. Because the AHE and the above-described SMR, which does not require spin polarization of the Pt, both generate the same electric response, it is difficult to disentangle the contribution of the two effects experimentally. Detailed ab-initio investigations of the magnitude of induced magnetization in \ce{Pt}, as well as the penetration depth, in \ce{(0001)Cr2O3-(111)Pt} heterostructures can help to reveal how much the AHE contributes to the Hall conductivity in Pt in magnetotransport experiments and/or spintronic devices.\\
\indent Although broadly accepted that \ce{Cr2O3} affects the electronic, and possibly magnetic (for example, spin-polarization of Pt via magnetic proximity to \ce{Cr2O3}) properties of \ce{Pt}, there has been little to no investigation of how the magnetic and electronic properties of \ce{Cr2O3} are altered due to proximity to \ce{Pt}. 
Elucidating whether \ce{Pt} alters \ce{(0001)Cr2O3} surface magnetization is important for establishing the degree to which the magnetic properties at \ce{Cr2O3} vacuum- and heavy metal-terminated surfaces can be directly compared.\\
\indent Based on these motivations, in this work we perform a detailed ab-initio investigation of the structural, electronic and magnetic properties at the \ce{(0001)Cr2O3-(111)Pt} interface. In Sec.~\ref{sec:structure} we calculate the energies of all distinct high-symmetry lateral alignments of \ce{(0001)Cr2O3} and \ce{(111)Pt} to determine the most energetically favorable lattice matching for the unreconstructed \ce{(0001)Cr2O3-(111)Pt} interface. In Sec. \ref{sec:electronic}, we analyze the charge density distribution and layer-projected density of states for the \ce{(0001)Cr2O3-(111)Pt} heterostructure with the lowest energy interface. We find a significant charge redistribution due to the interaction of \ce{Cr2O3} and \ce{Pt}, specifically, a substantial electron accumulation in the interstitial region between the \ce{Cr} and \ce{Pt} atoms closest to the interface. Finally, in Sec. \ref{sec:magnetic}, we shed light on the magnetic properties of both \ce{Cr2O3} and Pt near their interface, in particular, how the presence of Pt affects the \ce{Cr2O3} antiferromagnetic order (Subsec. \ref{subsec:Cr_flip}), and how the proximity of boundary magnetization in \ce{Cr2O3} affects the spin polarization of Pt (Subsec. \ref{subsec:MPE}).

\section{Computational methods\label{sec:methods}}
\begin{figure}
   \centering
   \includegraphics[]{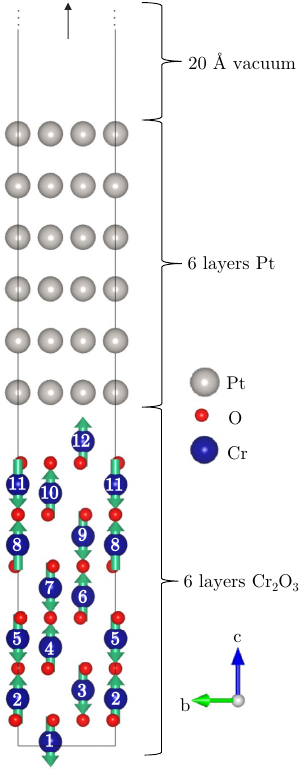}
   \caption{The \ce{Cr2O3-Pt} heterostructure consisting of six \ce{Cr2O3} layers, six Pt layers and 20 Å of vacuum. The antiferromagnetic structure is illustrated through the corresponding magnetic moments (green arrows). The numbers on the Cr atoms show the labels used in this work. Cr12 is the interfacial Cr atom.}
   \label{fig:input_structure}
\end{figure}

We use density functional theory (DFT) employing the Vienna ab-initio simulation package (VASP) \cite{kresse_efficient_1996, hafner_abinitio_2008} with projector-augmented-wave (PAW) pseudopotentials \cite{hafner_abinitio_2008, blochl_projector_1994}. The local spin-density approximation (LSDA) \cite{ziesche_density_1998, sousa_general_2007, csonka_assessing_2009, perdew_self-interaction_1981, ceperley_ground_1980} exchange-correlation functional is combined with a Hubbard U correction (LSDA+U) \cite{anisimov_band_1991} on the \ce{Cr} $d$ states in order to approximately account for their localized nature. We use the rotationally invariant method of Dudarev et al. \cite{dudarev_electron-energy-loss_1998} in which a single effective parameter $\mathrm{U}_{\mathrm{eff}}$ is adjusted, as opposed to separately specifying a Hubbard U and a Hund J. We set $\mathrm{U}_{\mathrm{eff}}=4$ $\mathrm{eV}$ based on experience from previous studies \cite{shi_magnetism_2009, weber_characterizing_2023}. The LSDA functional and the pseudopotentials $\mathrm{Cr_{sv}}$, $\mathrm{O}$, and $\mathrm{Pt_{pv}}$ with valence electrons of $\mathrm{Cr_{sv}}$:$3s^23p^63d^54s^1$; $\mathrm{O}$:$2s^22p^4$; and $\mathrm{Pt_{pv}}$:$5p^65d^96s^1$ are chosen based on good agreement of calculated and experimental lattice parameters, see appendix. Based on our convergence tests, we choose a 7x7x1 Gamma-centered k-point grid and a cutoff energy of 800 eV.\\
\indent For the calculations described in the main text, we use a slab structure with six layers of Cr-terminated \ce{(0001)Cr2O3}, six layers of \ce{(111)Pt}, and 20 Å vacuum, see Fig.~\ref{fig:input_structure}, unless specified otherwise. Pt forms a (111) surface on \ce{(0001)Cr2O3} as shown in several X-ray diffraction studies \cite{kilian_evidencing_2022, shimomura_enhancing_2017, moriyama_giant_2020, fiori_electronically_2023, iino_anomalous_2023, wang_increase_2022, nguyen_magnetic_2017}.
The structure files for \ce{Cr2O3} and Pt are obtained from the Materials Project database \cite{jain_commentary_2013} and transformed into a slab structure using the Atomic Simulation Environment (ASE) in python \cite{bahn_object-oriented_2002, hjorth_larsen_atomic_2017}. \ce{(111)Pt} with a lattice parameter of a=4.79 Å is strained to \ce{(0001)Cr2O3} with a lattice parameter of a=4.92 Å, yielding a lattice mismatch of $\approx 3 \%$. We fix the lattice parameters and relax the internal coordinates (while constraining the structure to maintain the original symmetry) until forces on all atoms are less than $0.01$ $\mathrm{eV}/\mathrm{\AA}$. During these relaxations, we fix the \ce{Cr} magnetic moments to the collinear ``up down up down" antiferromagnetic order with the N\'{e}el vector directed along $[0001]$ (Fig. \ref{fig:input_structure}), which is known to be the ground-state order for both bulk and vacuum-terminated \ce{(0001)Cr2O3} \cite{shi_magnetism_2009,weber_surface_2023}. Given the strong spin-orbit coupling of heavy metals such as Pt, we include spin-orbit coupling self-consistently for the heterostructure relaxations, as well as all subsequent calculations described in Secs.~\ref{sec:structure}, ~\ref{sec:electronic} and \ref{sec:magnetic}.

\section{Results}
\subsection{Structural properties\label{sec:structure}}
\begin{figure*}
    \centering
    \includegraphics{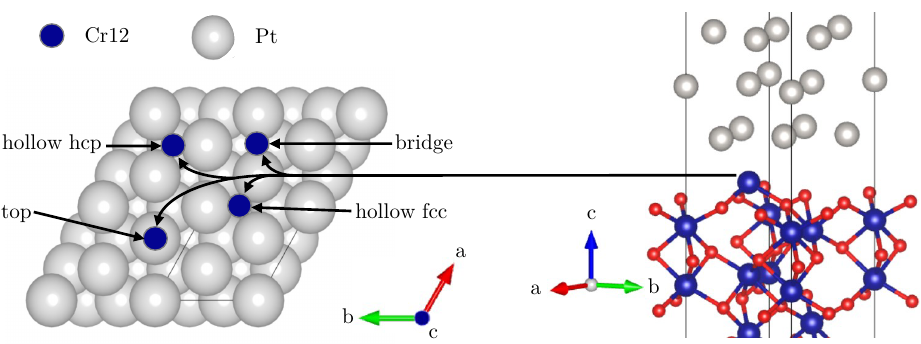}
    \caption{Terminology to describe the lateral alignment of the \ce{(0001)Cr2O3-(111)Pt} interface. Left: Top view of \ce{(111)Pt} surface, right: side view of hollow hcp (Pt on O) lateral alignment.}
    \label{fig:terminology}
\end{figure*}

First, we determine the most energetically favorable lateral alignment of the \ce{(0001)Cr2O3-(111)Pt} interface. An analogous ab-initio investigation of interface energetics in the isostructural heterostructure \ce{(0001)Fe2O3-(111)Pt} was performed in Ref. \cite{mahmoud_nature_2018}. For \ce{(0001)Cr2O3-(111)Pt} bilayers on the other hand, the one prior DFT calculation~\cite{moriyama_giant_2020} in the literature which we are aware of did not compare different lateral alignments.\\ 
\indent Possible adsorption sites for adatoms on the \ce{(111)Pt} surface are the so-called bridge, top, hollow fcc, and hollow hcp site \cite{zeng_density_2009} (Fig.~\ref{fig:terminology}). The difference between hollow fcc and hollow hcp is the presence of either a void or an atom in the second Pt layer beneath/above the adsorption site in the first Pt layer. We use this terminology to describe the lateral alignment of the interfacial Cr ion of \ce{(0001)Cr2O3} relative to the \ce{(111)Pt} surface in the \ce{Cr2O3-Pt} heterostructure (Fig.~\ref{fig:terminology}). Two possibilities arise for the hollow fcc and hcp sites: the Pt atoms of the first Pt layer can be positioned laterally above or between the O atoms of \ce{Cr2O3} resulting in the four configurations hollow fcc (Pt between O), hollow hcp (Pt between O), hollow fcc (Pt on O), and hollow hcp (Pt on O) in addition to the bridge and top configuration.\\
\indent We performed DFT calculations for these six lateral alignments to calculate the relaxed structure and determine the lowest energy interface configuration.
Note that we were not able to converge calculations using the bridge lateral alignment. The authors of Ref. \citenum{mahmoud_nature_2018} mentioned the same issue for their study of \ce{(0001)Fe2O3-(111)Pt} heterostructures, so we suspect that the bridge alignment is highly unstable. Tab.~\ref{tab:lateral_alignments_energies} gives the relative energies for the five remaining lateral alignments in $\mathrm{eV}$ per 6-layer-\ce{Cr2O3}-6-layer-Pt slab (48 atoms total) used in the calculations. Note that the atomic positions for each lateral alignment are fully relaxed as described in Sec. \ref{sec:methods} using the antiferromagnetic \ce{Cr2O3} order shown in Fig. \ref{fig:input_structure}. The effect of magnetic order will be studied later.
The hollow hcp (Pt on O) lateral alignment exhibits the lowest energy followed by the hollow fcc (Pt on O) lateral alignment. The other lateral alignments exhibit significantly higher energies. Thus, in addition to the clear energetic preference for Cr to lie between the Pt atoms in the interfacial layer, rather than directly below (corresponding to the top lateral alignment), it is also energetically favorable for Pt to be positioned on top of the oxygen atoms.

\begin{table}[h!tbp]
  \centering
  \caption{Energies of different lateral alignments relative to the hollow hcp (Pt on O) lowest-energy lateral alignment.}
  \label{tab:lateral_alignments_energies}
   \begin{tabular}{lc}
    \toprule
    Lateral alignment   & Energy difference [eV]  \\  
    \hline
    Top & +0.161  \\
    Hollow hcp (Pt between O)   & +0.158 \\
    Hollow fcc (Pt between O)   & +0.113  \\
    Hollow fcc (Pt on O) & +0.026  \\
    Hollow hcp (Pt on O) & 0.00  \\
    \hline
  \end{tabular}
\end{table}

The energy differences of the order of 0.01 to 0.1 eV per unit cell (Tab.\ref{tab:lateral_alignments_energies}) are significantly larger than the resolution of our calculations ($\approx$10$^{-6}$ eV), thus we can be confident that the hollow hcp (Pt on O) lateral alignment is the most energetically stable at 0 $\mathrm{K}$. Thus, for the remainder of the main text we focus on the electronic and magnetic properties of \ce{(0001)Cr2O3-(111)Pt} heterostructures with the hollow hcp (Pt on O) alignment (the results for other lateral alignments, given in the appendix, are qualitatively similar).\\
\indent Note that at room temperature (298 K), the thermal energy $k_B$T = 0.0257 $\mathrm{eV}$ is close to the energy difference between hollow hcp (Pt on O) and hollow fcc (Pt on O). Thus, at the growth temperature it is possible that the two lateral alignments coexist, or hollow fcc (Pt on O) could even dominate. However, at lower temperatures closer to the 0-kelvin DFT limit, we expect only the hollow hcp (Pt on O) alignment to be stable.\\
\indent Before moving on to the electronic properties, we stress that in this work, we focus on pristine, unreconstructed interfaces and have not considered possible surface reconstructions or sources of disorder such as vacancies or interstitials. Taking such disorder into account will be an important followup study. 

\subsection{Electronic properties\label{sec:electronic}}
We now analyze the electronic properties of the \ce{(0001)Cr2O3-(111)Pt} interface in the lowest-energy hollow hcp (Pt on O) lateral alignment. Specifically, we analyze the layer-projected density of states and calculate the real-space distribution of electronic charge using the charge density difference method.\\
\indent The charge density difference $\Delta\rho$ between the charge density of the \ce{Cr2O3-Pt} heterostructure $\rho_{\mathrm{Cr_2O_3-Pt}}$ and the charge density of a \ce{Cr2O3} slab $\rho_{\mathrm{Cr_2O_3}}$ and a Pt slab $\rho_{\mathrm{Pt}}$ can be expressed as 
\begin{equation}
  \Delta\rho = \rho_{\mathrm{Cr_2O_3-Pt}} - \rho_{\mathrm{Cr_2O_3}} - \rho_{\mathrm{Pt}}.
  \label{eq:CCD}
\end{equation}

\begin{figure*}
    \centering
    \includegraphics{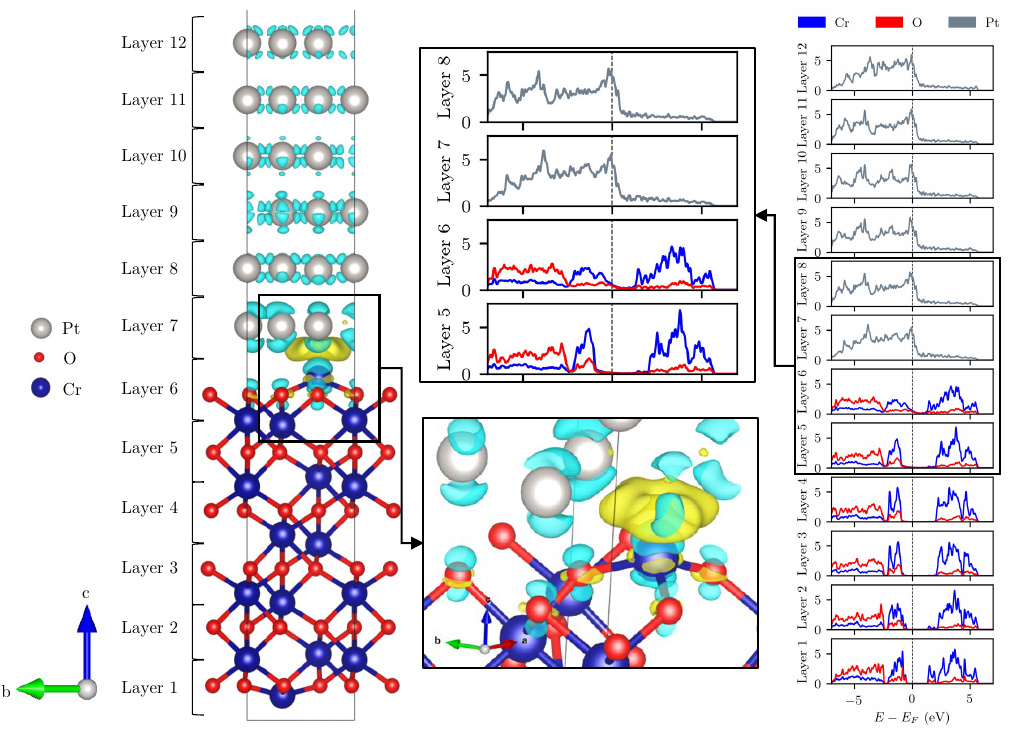}
    \caption{Left: Calculated charge density difference for the hollow hcp (Pt on O) lateral alignment \ce{(0001)Cr2O3-(111)Pt} heterostructure. Yellow represents an accumulation of electrons, while cyan represents a depletion of electrons. The isosurface value is set to 0.006 $\frac{e}{\si{\cubic\angstrom}}$. Right: Layer-projected density of states, with the layers defined in the left-hand structure. Insets in the center zoom in on properties near the interface.}
    \label{fig:electronic_properties} 
\end{figure*}

$\Delta\rho$ captures the changes in electron density, relative to the summed densities of isolated \ce{Cr2O3} and Pt structures, due to the mutual interaction at their interface, and is visualized with the structure visualization software VESTA in Fig. \ref{fig:electronic_properties}. While no charge transfer is visible for layers 1 to 5 in \ce{Cr2O3}, we observe a charge density difference in all Pt layers as well as in the layer of \ce{Cr2O3} at the \ce{Cr2O3-Pt} interface. The Pt atoms exhibit a slight depletion of electrons throughout all Pt layers, marked in blue in Fig. \ref{fig:electronic_properties}. Electrons accumulate between the Cr12 interface ion and the first Pt layer. Note that this accumulation of electrons, depicted in yellow, is present for all lateral alignments, with only the shape of the electron accumulation changing, see appendix.

The accumulation of electrons at the interface indicates a metallic behavior that is also reflected in the layer-projected density of states, shown on the right in Fig. \ref{fig:electronic_properties}. While layers 1 to 5 in \ce{Cr2O3} are insulating with a pronounced band gap, the interface layer (layer 6) shows a finite density of states at the Fermi energy and no clear bandgap.
These effects are also robust across different alignments, see appendix.\\
\indent The substantial charge accumulation between the Cr12 ions closest to the interface and the Pt directly above already provides a qualitative hint that the properties of Cr moments at the \ce{(001)Cr2O3-(111)Pt} interface may be altered relative to bulk or vacuum-terminated \ce{Cr2O3}. This is especially evident in the \ce{Cr2O3} magnetism, as we will see in the  next section.

\subsection{Magnetic properties\label{sec:magnetic}}
Next, we calculate the magnetic properties of the \ce{(0001)Cr2O3-(111)Pt} interface to investigate whether the surface magnetization of \ce{Cr2O3} is affected by Pt, and whether Pt is affected by \ce{Cr2O3}.

\subsubsection{Effect of Pt on \ce{Cr2O3} -- Flipped magnetic moment\label{subsec:Cr_flip}}
\begin{figure}
    \centering
    \includegraphics[]{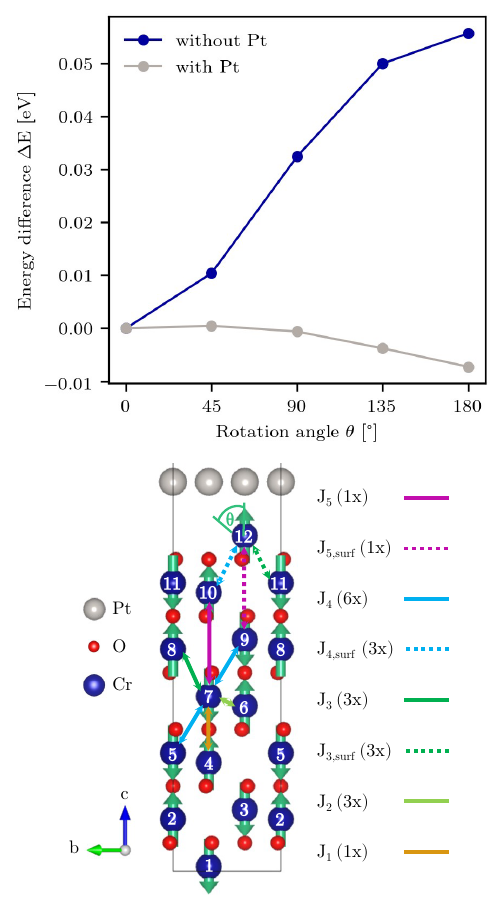}
    \caption{Top: Energies calculated at different rotation angles (with $\theta$ defined in the bottom figure) of the Cr12 interface magnetic moment relative to the energy at 0 degrees. 0 degrees corresponds to the antiferromagnetic structure of \ce{Cr2O3}, while 180 degrees corresponds to Cr12 reversed (pointing downwards in this case). Bottom: Exchange interaction coupling parameters, described further in the text, for both bulk-like (Cr7) and surface/interface (Cr12) magnetic sites.}
    \label{fig:magnetic_properties_Cr2O3}
\end{figure}

First, we determine how the presence of Pt in the \ce{(0001)Cr2O3-(111)Pt} heterostructure affects the \ce{(0001)Cr2O3} surface magnetization, in particular the monolayer of Cr12 moments closest to the vacuum or Pt interface. By revealing how the presence of the heavy metal modifies \ce{Cr2O3} magnetic properties at the interface compared to the vacuum-terminated surface, our investigation should indicate how directly magnetotransport measurements exploiting electrical readout in heavy metals can be compared with alternative experimental surface magnetization probes, such as nitrogen vacancy magnetometry.\cite{appel_nanomagnetism_2019,makushko_flexomagnetism_2022}.\\
\indent We use constrained magnetic DFT as implemented by Ma and Dudarev \cite{ma_constrained_2015} to rotate the Cr12 interface magnetic moment in steps of 45$^\circ$ from the bulk orientation, in which it is antiferromagnetically aligned with its nearest neighbor Cr11, to the opposite orientation. We then calculate the corresponding energies to check whether canting of the surface magnetization is energetically favorable. Without Pt (vacuum termination), the energy is lowest at 0$^\circ$, that is, when the \ce{Cr2O3} slab retains its bulk antiferromagnetic ordering (Fig.~\ref{fig:magnetic_properties_Cr2O3}). This is consistent with our prior DFT studies using vacuum-terminated \ce{(0001)Cr2O3}~\cite{weber_characterizing_2023} as well as ab-initio studies by other authors~\cite{wysocki_microscopic_2012}.\\
\indent In stark contrast, for the \ce{(0001)Cr2O3-(111)Pt} heterostructure, the energy is lowest when the magnetic moment of Cr12 is rotated by 180$^\circ$ (Fig. \ref{fig:magnetic_properties_Cr2O3}). This has the surprising implication that, at least for pristine, unreconstructed \ce{(0001)Cr2O3-(111)Pt} heterostructures, the interface magnetic moment of Cr12 flips with respect to its direction in bulk and vacuum-terminated \ce{(0001)Cr2O3}, disrupting the antiferromagnetic order at the interface and leading to a net magnetic moment in the \ce{Cr2O3} slab. We also confirm that the flipped Cr interface magnetic moment is energetically favorable for the four additional lateral alignments we explored in Sec. \ref{sec:structure}, see appendix. Thus, we can conclude that the reversal of \ce{(0001)Cr2O3} magnetization at the interface is a generic consequence of \ce{Cr2O3-Pt} interactions, and is robust to the microscopic details of different lateral alignments. This is a key result of this work.\\
\textbf{Exchange interactions:} Next, we calculate the isotropic Heisenberg spin exchange interactions for Cr12 moments in the \ce{(0001)Cr2O3-(111)Pt} heterostructure and compare them to those in bulk \ce{Cr2O3} in order to rationalize the flipping of the interface moments. We take the convention with the Heisenberg contribution to the energy given by
\begin{equation}
    H_{\mathrm{Heis}}=\sum_{\left<i,j\right>}J_{ij}\mathbf{\hat{e}}_i\cdot \mathbf{\hat{e}}_j,
    \label{eq:Hheis}
\end{equation}
where $J_{ij}$ is the Heisenberg exchange coupling parameter between magnetic moments on sites $i$ and $j$, and $\mathbf{\hat{e}}_i$ is the unit vector parallel to the magnetic moment of the Cr ion at site $i$. Note that the convention of Eq. \ref{eq:Hheis} absorbs the value of spin into the exchange constant, and thus when comparing this work's $J_{ij}$ values to literature values where the Hamiltonian is written as $H_{\mathrm{Heis}}=\sum_{\left<i,j\right>}J_{ij}\mathbf{S}_i\cdot \mathbf{S}_j$, such as Ref. \citenum{shi_magnetism_2009}, it is necessary to divide by $S^2$ with $S$ being the spin of the \ce{Cr^{3+}} ion. Also, with the convention of Eq. \ref{eq:Hheis}, positive values of $J_{ij}$ indicate preferred antiferromagnetic alignment for moments $i$ and $j$, whereas negative values indicate preferred ferromagnetic alignment. We use the four-state total energy method~\cite{xiang_predicting_2011} whereby the exchange coupling parameter $J_{ij}$ of the magnetic moments on atoms $i$ and $j$ is calculated by constraining the magnetic moments of atoms $i$ and $j$ to all four possible collinear configurations $\uparrow\uparrow$, $\downarrow\downarrow$, $\downarrow\uparrow$, and $\uparrow\downarrow$ while all other magnetic moments remain the same. Then

\begin{equation}
    J_{ij} = \frac{E(\uparrow\uparrow)+E(\downarrow\downarrow)-E(\downarrow\uparrow)-E(\uparrow\downarrow)}{4 N},
    \label{eq:4state}
\end{equation}
with $N$ being the coupling degeneracy (the number of symmetry-equivalent nearest neighbors corresponding to the $J_{ij}$ of interest)  and $E$ the total DFT energy for each $(ij)$ configuration~\cite{shi_magnetism_2009, xiang_predicting_2011}.\\
\indent There are five relevant exchange interaction coupling parameters $J_1$ to $J_5$ for bulk \ce{Cr2O3},with $J_n$ denoting the $n$th nearest neighbor. Couplings beyond $J_5$ can be safely neglected \cite{shi_magnetism_2009}. At the Cr-terminated surface of \ce{Cr2O3}, the dominating, antiferromagnetic $J_1$ and $J_2$ interactions are cut off and only $J_{3,surf}$ to $J_{5,surf}$ are retained, as illustrated in Fig.~\ref{fig:magnetic_properties_Cr2O3}. As a result, Cr12 in vacuum-terminated \ce{(0001)Cr2O3} is only weakly coupled to other Cr moments, and consequently the surface magnetization is effectively paramagnetic at room temperature, even though the bulk N\'{e}el vector is magnetically ordered~\cite{weber_characterizing_2023}.\\
\indent We first calculate the Heisenberg exchange parameters for a Cr site in the center of the \ce{Cr2O3} slab (specifically Cr7, Fig. \ref{fig:magnetic_properties_Cr2O3}) to see whether Pt affects the magnetic structure far from the interface. The values (we show only $J_1$ and $J_2$ in Tab.~\ref{tab:J_values} as these are the dominant couplings in bulk) are very close to our previous findings in bulk \ce{Cr2O3} see Tab. \ref{tab:J_values}. Consequently, we can be confident that the heterostructure we use in our DFT calculations is sufficiently thick to reproduce bulk-like behavior of \ce{Cr2O3} in the middle of the slab (this is also reflected in the layer-projected density of states in Fig. \ref{fig:electronic_properties} in Sec. \ref{sec:electronic}).\\
\indent In contrast, the surface exchange parameters for the interface Cr12 are strongly affected by the presence of Pt. The most significant change with respect to the values in a relaxed vacuum-terminated slab of \ce{(0001)Cr2O3} (Tab.~\ref{tab:J_values}, Ref.~\cite{weber_characterizing_2023}) is for $J_{3,surf}$, which increases by more than a factor of $20$ in the heterostructure and becomes ferromagnetic. This indicates a preference for Cr12 and the Cr11 just below to align parallel, consistent with the observed energy lowering in Fig. \ref{fig:magnetic_properties_Cr2O3} for flipping Cr12 with respect to the antiferromagnetic order. $J_{4,surf}$ ($J_{5,surf}$) on the other hand favor ferromagnetic (antiferromagnetic) alignment between Cr12 and Cr10 (Cr 9), both of which favor the normal orientation $\theta=0$ for Cr12. However, by calculating the contribution to the Heisenberg energy if Cr12 is flipped, based on Eq.~\ref{eq:Hheis} and the coupling degeneracies in Tab.~\ref{tab:J_values}, we obtain 
\begin{multline}
    H_{\mathrm{Heis}}^{\mathrm{Cr12,flipped}}=3J_{3,surf}-3J_{4,surf}+J_{5,surf}\\
    =-2.2\mathrm{meV}.
    \label{eq:flippedE}
\end{multline}
Thus, the flipping of the top layer next to the \ce{(0001)Cr2O3-(111)Pt} interfaces is overall energetically favorable due to the strong ferromagnetic $J_{3,surf}$ in the presence of Pt \footnote{We must mention that the rigorous quantitative accuracy of $J_{ij}$ as obtained from Eq. \ref{eq:4state} relies on the magnetic and electric properties of all atoms other than $i$ and $j$ remaining unchanged for the calculations. If this is the case, then $E(\theta=0)-E(\theta=180^{\circ})$, which as shown in Fig. \ref{fig:magnetic_properties_Cr2O3} is about 7 $\mathrm{meV}$, should be precisely equal to $H_{\mathrm{Heis}}^{\mathrm{Cr12},\theta=0}-H_{\mathrm{Heis}}^{\mathrm{Cr12},\theta=180^{\circ}}$. However, using the values in Tab.\ref{tab:J_values}, we find that the difference between flipped and antiferromagnetic configurations of Cr12 is 4.4 $\mathrm{meV}$. The slight difference comes from the fact that in some of the configurations used in the 4-state method, specifically $E(\downarrow \uparrow)$, the Pt magnetization reverses (discussed in Sec. \ref{subsec:MPE}) with respect to its orientation in the three other magnetic configurations. Thus, our $J$ values include a small contribution due to \ce{Cr2O3-Pt} interactions which does not cancel. However, the \emph{relative} values of $J_{3,surf}$, $J_{4,surf}$, and $J_{5,surf}$, which are what determines the energetic favorability of the flipped \ce{Cr2O3} configuration in the presence of Pt, should be reliable.}.\\
\indent Note however, that the magnitude $\sim 2$ $\mathrm{meV}$ of the effective coupling for $Cr12$ in the \ce{(0001)Cr2O3-(111)Pt} bilayer is very similar to the magnitude of effective coupling for Cr12 with vacuum termination\cite{weber_characterizing_2023}. Thus, we would expect Cr12 to also be effectively paramagnetic near the bulk N\'{e}el temperature in the presence of Pt. The key difference is that at low temperature where Cr12 is ordered, the interaction with Pt flips the effective coupling, and hence the ground-state direction of Cr12, with respect to the vacuum-terminated case.
\begin{table}[h!tbp]
 \centering 
 \caption{Exchange parameters $J_n$ in bulk \ce{Cr2O3}, the \ce{(0001)Cr2O3-(111)Pt} bilayer with hollow hcp (Pt on O) lateral alignment, and a relaxed vacuum-terminated \ce{(0001)Cr2O3} slab as calculated in Ref.~\cite{weber_characterizing_2023}. $N$ indicates the coupling degeneracy. Negative values of $J$ indicate ferromagnetic couplings, and positive values refer to antiferromagnetic couplings. (Note that in Ref. \cite{weber_characterizing_2023} we used a different convention for the Heisenberg Hamiltonian such that all reported values were divided by two and multiplied by minus one relative to this table.)} 
 \label{tab:J_values}
\begin{tabular}{l c c c c}
\toprule
 &   &bulk \ce{Cr2O3}~\cite{weber_characterizing_2023} & \ce{Cr2O3-Pt}    & \ce{Cr2O3}-vac.~\cite{weber_characterizing_2023} \\
$J_n$ & $N$& $J$ [meV] & $J$ [meV] & $J$ [meV] \\
\hline    
$J_1$       & 1 & 20.92 & 21.10 & $-$                                               \\
$J_2$       & 3 & 15.76 & 15.98 & $-$                                              \\
$J_{3,surf}$ & 3 & $-$ & -10.50 & 0.30                                                 \\
$J_{4,surf}$ & 3 & $-$ & -9.01 & -8.88                                              \\
$J_{5,surf}$ & 1 & $-$ & 2.20 & 0.78                                               \\
\bottomrule
\end{tabular}
\end{table}

\subsubsection{Effect of \ce{Cr2O3} on Pt -- Magnetic proximity effect\label{subsec:MPE}}
\begin{figure}
    \centering
    \includegraphics[width=0.99\columnwidth]{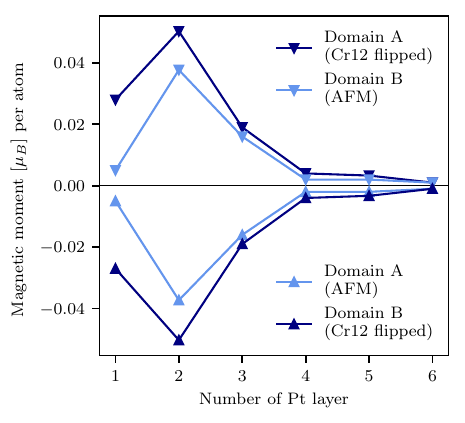}
    \caption{Magnetization in Pt for the two different domain states in \ce{Cr2O3} for the antiferromagnetic configuration and the configuration with Cr12 flipped interface magnetic moment. Layer 1 is located towards \ce{Cr2O3} and layer 6 towards vacuum. A magnetic proximity effect is present in the first three Pt layers and changes sign when the Cr12 magnetic moment is flipped.}
    \label{fig:MPE}
\end{figure}

Finally, we briefly investigate the influence of antiferromagnetic \ce{Cr2O3} on Pt. A spin polarization of Pt induced via proximity to \ce{Cr2O3}, that is a magnetic proximity effect is a-priori likely, considering that Pt is close to the Stoner criterion.  Our DFT calculations reveal a sizeable magnetization of the order of $\pm$ 0.01 $\mu_B$ per atom in the first three Pt atomic layers at the interface with \ce{Cr2O3} (Fig. \ref{fig:MPE}). Although Pt is strained to \ce{Cr2O3}, magnetization caused by strain can be excluded, as a calculation of a Pt slab strained to match the \ce{Cr2O3} lattice constant gives zero magnetization in Pt without the presence of \ce{Cr2O3}. From the fourth layer onward, the magnetization becomes negligible, dropping down to the order of $\pm$ 0.001 $\mu_B$. We find that the Pt magnetic polarization is confined to the first three layers at the interface in all other lateral alignments as well, see appendix. Thus, based on our calculations, Pt in \ce{(0001)Cr2O3-(111)Pt} heterostructures exhibits a clear and localized magnetic proximity effect (MPE).\\
\indent Moreover, the magnetization in Pt is inherently coupled to the antiferromagnetic domain state of \ce{Cr2O3}. When the domain state of \ce{Cr2O3} is switched (Domain A $\rightarrow$ Domain B), the magnetization values in Pt have the same magnitude but the opposite sign (Fig. \ref{fig:MPE}). The coupling of Pt to the Cr interface magnetic moment, Cr12, is antiferromagnetic for all three Pt layers with significant spin polarization with, interestingly, the maximum spin polarization occurring in the second Pt layer from the \ce{Cr2O3-Pt} interface, rather than in the Pt layer directly adjacent to \ce{Cr2O3}.\\
\indent Lastly, related to the prior section \ref{subsec:Cr_flip} in which we found that for a given domain, the interface Cr12 moments flip with respect to their standard antiferromagnetic order, we also investigate the effect on spin polarization in Pt when Cr12 has the flipped $(\theta=180^{\circ})$, rather than antiferromagnetic $(\theta=0^{\circ})$ orientation. Importantly for experimental implications, from the dark blue lines in Fig. \ref{fig:MPE} we see that flipping the Cr12 magnetic moment into its lowest energy configuration has a similar effect on Pt as switching the domain state of \ce{Cr2O3}. In addition, the \ce{Cr2O3-Pt} interface with flipped Cr interface magnetic moment has an enhanced magnetic proximity effect in the first two Pt layers compared to the \ce{Cr2O3-Pt} interface with antiferromagnetic order in \ce{Cr2O3}.


\section{Discussion and Conclusions}
In summary, we have carried out an in-depth computational exploration of the interface energetics, and the electronic and magnetic properties of \ce{(0001)Cr2O3-(111)Pt} heterostructures.

Our analysis of all possible pristine lateral alignments shows that interface stackings (both fcc and hcp) with Pt directly on top of O  are both very close in energy to each other, as well as being substantially lower in energy than all other lateral alignments. This finding indicates that the position of Pt relative to O is more relevant in the heterostructure energetics than whether the alignment is hollow hcp or hollow fcc. We also stress again, that the experimental relevance of our findings on lateral alignment stability must be explored further in followup studies by taking into account likely possibilities of surface reconstruction and other disorder.\\
\indent Our calculations for the lowest-energy hollow hcp (Pt on O) alignment indicate substantial electronic and magnetic effects due to \ce{Cr2O3-Pt} interactions in the layers closest to the interface. Beyond the interface region of one \ce{Cr2O3} layer and three Pt layers, both Pt and \ce{Cr2O3} are bulklike.\\
\indent One of our most potentially consequential findings is that a reversed Cr interface magnetic moment in \ce{(0001)Cr2O3-(111)Pt heterostructures} is energetically more favorable than the usual antiferromagnetic structure of \ce{Cr2O3}. This is reflected also in our calculation of the Heisenberg coupling constants for the interface Cr, in particular, a strongly ferromagnetic $J_{3,surf}$ which favors a flipping of the interface Cr compared to its alignment in the bulk or with vacuum termination. The microscopic mechanism of this flipping needs to be investigated further. Qualitatively, it is presumably tied to the observed accumulation of charge density between Cr and Pt interface moments.\\ 
\indent Our finding of the flipping of interface Cr for \ce{(0001)Cr2O3-(111)Pt} heterostructures suggests that heavy metals cannot be assumed to be passive readout elements for surface magnetic properties, but rather, their presence can drastically alter the magnetism at the surface. For \ce{(0001)Cr2O3-(111)Pt} heterostructures in particular, our findings (Fig.~\ref{fig:MPE}) reveal that a given antiferromagnetic domain with the vacuum-terminated \ce{(0001)Cr2O3} surface moment orientation produces a qualitatively identical response in Pt as the opposite \ce{(0001)Cr2O3} domain with the flipped interface moment orientation. Thus, it is important to verify whether for example established field-cooling procedures are actually switching bulk \ce{Cr2O3} domains, or simply switching the interface \ce{Cr} with a fixed bulk domain.\\
 \indent Additionally, we find a substantial magnetization $\sim0.01\mu_B$ in the first three layers of Pt induced via \ce{Cr2O3}. This calculated magnetic proximity effect is consistent with experiments utilizing the magneto-optical Kerr effect (MOKE) measurements~\cite{cao_magnetization_2017}, though it must be emphasized that spurious signals in Kerr and soft x-ray measurements can arise due to dirty samples and be confused with magnetic signals from a pure sample. In contrast, other measurements using x-ray magnetic circular dichrosim (XMCD) do not find a significant MPE in \ce{(0001)Cr2O3-(111)Pt} heterostructures \cite{moriyama_giant_2020}. We hope our work motivates experimentalists to look more closely into the possible spin polarization of Pt, which via breaking of time-reversal symmetry would allow the AHE in Pt on \ce{(0001)Cr2O3}.\\
 \indent If an MPE in Pt for \ce{(0001)Cr2O3-(111)Pt} exists, an interesting experiment to test whether it contributes substantial at low temperatures to the transverse voltage signal is to compare Pt electrical resistance at depths within and beyond three layers from the interface. Based on the apparent penetration depth of the MPE, the electrical response within the first three layers of Pt should contain an additional component due to the AHE induced by the magnetized Pt. This will disappear in deeper regions of the Pt as the MPE falls off.\\
\indent To conclude, our first-principles investigation of structural, electronic and magnetic properties of \ce{(0001)Cr2O3-(111)Pt} heterostructures can help with interpretation of magnetotransport experiments in assessing the sign of the N\'{e}el vector, as well as provide guidance in the future design of AFM spintronic devices. We also hope that our work inspires further investigation, both theoretical and experimental, of how the interactions of antiferromagnetic and heavy metal materials can lead to unexpected electric and magnetic phenomena at their interfaces. 


\section*{Acknowledgments}
We thank Denys Makarov and Oleksandr Pylypovskyi for useful discussions. This work was funded by the ERC under the European Union’s Horizon 2020 research and innovation programme with grant No. 810451, and by ETH Z\"urich. Computational resources were provided by ETH Z\"urich’s EULER cluster. 
\section*{Appendix}
\appendix
\section{DFT Details: Assessment of functionals}
We relaxed the lattice parameters of bulk \ce{Pt} and bulk \ce{Cr2O3} with different DFT functionals and compared them to experimental values to determine the functional that best describes the \ce{Cr2O3-Pt} heterostructure.
Tables \ref{tab:Cr2O3_functional_comparison} and \ref{tab:Pt_functional_comparison} give the values of the lattice parameters of \ce{Cr2O3} and Pt obtained with the local (spin) density approximation (L(S)DA) and the Perdew-Burke-Ernzerhof (PBE) functional. A Hubbard U correction is applied on the Cr 3d orbitals in both cases, as described in Sec. \ref{sec:methods}.
We used the VASP Cr\_sv and O PAW pseudopotentials as established in previous studies \cite{weber_characterizing_2023}, and compared the standard pseudopotential (Pt) and extended pseudopotential (Pt\_pv) for platinum for the LDA and the PBE functional. The extended pseudopotentials ending with \_pv and \_sv additionally take into account semi-core p and s electrons respectively as valence electrons in addition to the usual outer shell electrons.

\begin{table}[h!tbp]
  \centering
  \caption[Comparison of functionals for \ce{Cr2O3}]{Relaxed lattice parameters of \ce{Cr2O3} computationally obtained with LDA and PBE functionals compared to the experimental value determined in Ref. \cite{hill_crystallographic_2010}. $\Delta$a and $\Delta$c describe the relative difference between the computational and experimental lattice parameters.}
  \label{tab:Cr2O3_functional_comparison}
  \begin{tabular}{lcccc}
    \toprule
     & a [Å] & $\Delta$a & c [Å] & $\Delta$c \\
    \hline
    LSDA & 4.9156 & --0.84 \% & 13.5186  &  --0.54 \% \\
    PBE & 5.0340 & +1.55 \% & 13.8026 & +1.55 \% \\
    Experiment \cite{hill_crystallographic_2010} & 4.9572 & -- & 13.5917 & -- \\
    \bottomrule
  \end{tabular}
\end{table}

\begin{table}[h!tbp]
  \centering
  \caption[Comparison of functionals for Pt]{Relaxed lattice parameters of \ce{Pt} computationally obtained with LDA and PBE functionals in combination with Pt and Pt\_pv PAW pseudopotentials compared to the experimental value determined in Ref. \cite{arblaster_crystallographic_1997}. $\Delta$a describes the relative difference between the computational and experimental lattice parameters.}
  \label{tab:Pt_functional_comparison}
  \begin{tabular}{lcc}
    \toprule
     & a [Å] & $\Delta$a \\  
    \hline
    LDA (Pt) & 3.8326 & --2.3 \% \\
    LDA (Pt\_pv) & 3.9096  & --0.4 \% \\
    PBE (Pt) & 3.9137 & --0.3 \% \\
    PBE (Pt\_pv) & 3.9771 &  +1.4\% \\
    Experiment \cite{arblaster_crystallographic_1997} & 3.9236  & -- \\
    \bottomrule
  \end{tabular}
\end{table}

The results in Tables \ref{tab:Cr2O3_functional_comparison} and \ref{tab:Pt_functional_comparison} show that the LDA functional with the extended Pt\_pv pseudopotential yields lattice parameters closest to the experimental lattice parameters. Based on these results, the L(S)DA functional (+U correction) as well as Cr\_sv, O, and Pt\_pv pseudopotentials are selected to model the \ce{Cr2O3-Pt} heterostructure.

\section{Electronic properties - More lateral alignments}
\label{sec:appendix_electronic}
\begin{figure}
    \centering
    \includegraphics[]{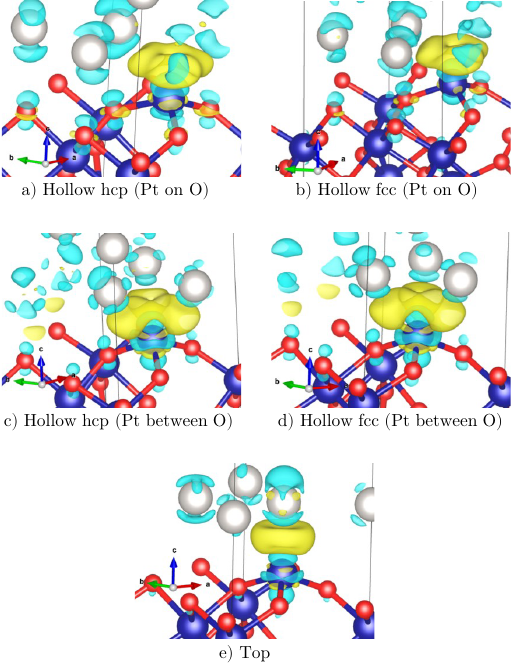}
    \caption{Charge density differences at the interface between Cr12 and the first Pt layer for \ce{(0001)Cr2O3-(111)Pt} heterostructures with different lateral alignments. Yellow represents an accumulation of electrons, while cyan represents a depletion of electrons The isosurface value in VESTA is set to 0.006 $\frac{e}{\si{\cubic\angstrom}}$.}
    \label{fig:blobs}
\end{figure}

While Sec. \ref{sec:electronic} focuses on the electronic properties of the lowest-energy lateral alignment, here we present the electronic properties of the other lateral alignments. We find an electron accumulation in the \ce{Cr2O3} interface layer for all lateral alignments, with the detailed shape of the electron accumulation depending on the lateral alignment, as shown in Fig. \ref{fig:blobs}.
The hollow fcc and hollow hcp configurations have a triangular-shaped accumulation of electrons, with the triangle corners pointing towards the Pt atoms. 
The hollow fcc (Pt between O) and hollow hcp (Pt between O) lateral alignment show similar accumulations to each other; likewise the hollow fcc (Pt on O) and hollow hcp (Pt on O) alignments are similar. Consequently, the relative \ce{O-Pt} position plays a larger role in the shape of the electron accumulation than whether the lateral alignment is of fcc or hcp type. 
The electron accumulation in the `top' lateral alignment occurs in a doughnut-like shape between the Pt atom above and the Cr12 ion below.
The shape of the electron accumulation stays similar upon variation of the isosurface value for all different lateral alignments.
We note that the layer-projected density of states also shows a finite DOS in the \ce{Cr2O3} interface layer for all lateral alignments, consistent with the observed electron accumulations in Fig. \ref{fig:blobs}.

Overall, the electron accumulations appear in proximity to the Cr12 ion irrespective of whether the Cr12 magnetic moment is flipped or not, and their shape is controlled by the platinum interface lateral alignment.

\section{Magnetic properties - Thicker slabs and supercell}
In the main text, \ce{Cr2O3-Pt} heterostructures with six \ce{Cr2O3} layers are evaluated. In this section, the effect of an increasing number of \ce{Cr2O3} layers is analyzed.

The resulting energy differences between antiferromagnetic order and magnetic order with Cr12 flipped magnetic moment of four slab structures with different numbers of \ce{Cr2O3} layers are listed in Table \ref{tab:energy_differences_AFM_Cr12flipped}. The flipped Cr12 magnetic moment is energetically favorable for all four of them. The energy difference is similar for all four, and importantly there is no trend in increasing or decreasing energy difference that would suggest an additional thickness-dependent contribution to the energy. 

\begin{table}[h!tbp]
  \centering
  \caption{Energy difference $\Delta$E between the flipped and antiferromagnetic Cr12 configurations for different slabs of the lowest energy lateral alignment (hollow hcp (Pt on O)) with four Pt layers.}
  \label{tab:energy_differences_AFM_Cr12flipped}
\begin{tabular}{@{}cc@{}}
\toprule
No. \ce{Cr2O3} layers    & Energy difference $\Delta$E [meV] \\
\hline
6         &  -18.3\\
7   &  -15.0\\
9    & -16.1 \\
12 & -19.1 \\
\bottomrule
\end{tabular}
\end{table}

In addition, we analyzed the Cr interface magnetic moments in a 2x1x1 supercell to check whether a combination of flipped and AFM interface $\mathrm{Cr}$ atoms is lower in energy than flipping all $\mathrm{Cr}$ moments. We found that the 2x1x1 supercell has lowest energy for both top Cr12 flipped, intermediate energy for one Cr12 flipped and one with the AFM configuration, and highest energy for both primitive magnetic cells having the bulk AFM order for the Cr12 moments. Thus, it seems that lateral spin rearrangements of \ce{Cr} at the \ce{(0001)Cr2O3-(111)Pt} which lead to a mixed flipped-AFM interface do not lower the energy compared to a uniformly flipped Cr12 layer. Moreover, the energy difference between both Cr flipped and both Cr in the bulk AFM configuration for the 2x1x1 supercell is almost exactly twice as large as the energy difference for the single magnetic unit cell used in the main text. This implies that no other energy contributions (for example, a non-negligible magnetic exchange between adjacent Cr12 moments mediated via RKKY interactions) which were absent in the primitive magnetic cell arise in the supercell containing two Cr12 moments. Thus, interactions between neighboring surface Cr moments in the heterostructure, like the case of the vacuum-terminated structure, can be safely neglected, and the observed flipping of Cr surface moments for the single magnetic unit cell should occur for all Cr12 moments at an experimental \ce{(0001)Cr2O3-(111)Pt} interface.

\section{Magnetic properties - Further analysis of MPE}
Section \ref{subsec:MPE} described the observation of a magnetic proximity effect (MPE) for a \ce{(0001)Cr2O3-(111)Pt} heterostructure with six Pt layers in a hollow hcp (Pt on O) lateral alignment. This section analyzes the MPE of the other lateral alignments as well as its dependence on the thickness of Pt.

\begin{figure}
    \centering
    \includegraphics[]{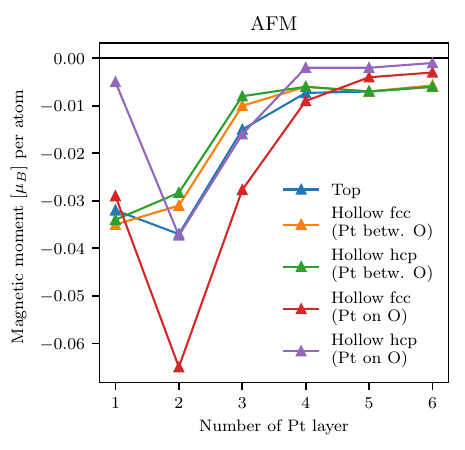}
    \caption{Magnetization of Pt atoms in the hexagonal [0001] direction as a function of Pt layer for all different lateral alignments with antiferromagnetic order in \ce{Cr2O3}. A magnetic proximity effect is pronounced in the first three Pt layers.}
    \label{fig:MPE_all_lateral_alignments_AFM}
\end{figure}

\begin{figure}
    \centering
    \includegraphics[]{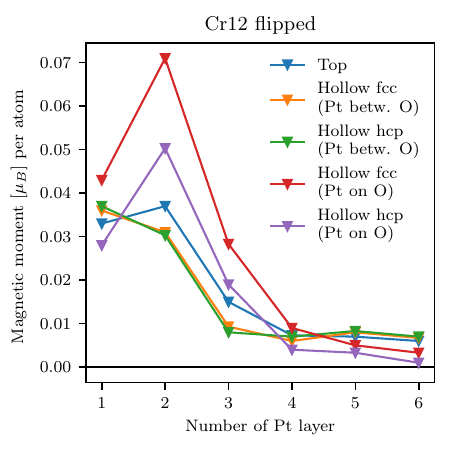}
    \caption{Magnetization of Pt atoms in the hexagonal [0001] direction as a function of Pt layer number for different lateral alignments. The \ce{Cr2O3} slab has the Cr12 magnetic moment flipped, in agreement with our calculated ground state for the heterostructure. A magnetic proximity effect is pronounced in the first three Pt layers.}
    \label{fig:MPE_all_lateral_alignments_Cr12flipped}
\end{figure}

Fig. \ref{fig:MPE_all_lateral_alignments_AFM} and Fig. \ref{fig:MPE_all_lateral_alignments_Cr12flipped} show the calculated magnetic moment per Pt atom as a function of the Pt layer number for different lateral alignments. For all cases, the Pt magnetization points along the [0001] hexagonal direction, perpendicular to the \ce{(0001)Cr2O3-(111)Pt} interface, with negligible in-plane components. We see that the MPE is present for all lateral alignments, and in all cases occurs in the first three Pt layers adjacent to \ce{Cr2O3}. The hollow fcc (Pt on O) magnetization values (purple) and hollow hcp (Pt on O) magnetization values (brown) have the same dependence on distance from the interface; likewise, the hollow fcc (Pt between O) values (green) and hollow hcp (Pt between O) values (red) have the same dependence. This suggests that the relative alignment of Pt and O affects the MPE more than whether the alignment is of fcc or hcp type. This is analogous to the behavior of the electron accumulation described in Sec. \ref{sec:appendix_electronic}.

\begin{figure}
    \centering
    \includegraphics[]{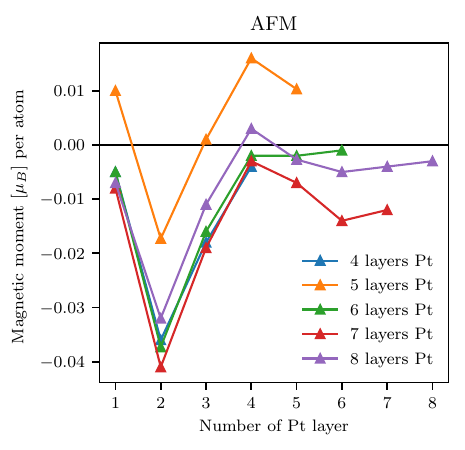}
    \caption{Magnetization of Pt atoms in the hexagonal [0001] direction as a function of Pt layer number for different numbers of Pt layer thicknesses with antiferromagnetic order in \ce{Cr2O3}. A magnetic proximity effect is pronounced in the first three Pt layers.}
    \label{fig:MPE_number_of_layers_AFM}
\end{figure}

\begin{figure}
    \centering
    \includegraphics[]{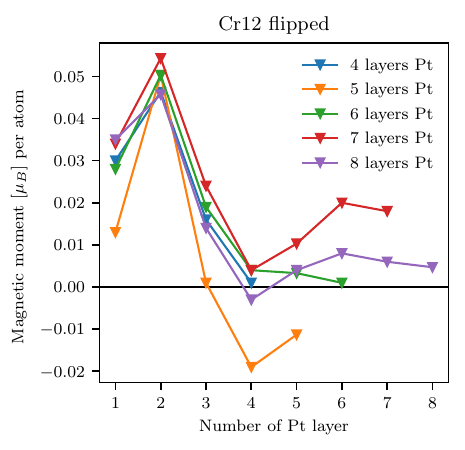}
    \caption{Magnetization of Pt atoms in the hexagonal [0001] direction as a function of Pt layer number for different Pt layer thicknesses with flipped Cr12 magnetic moment in \ce{Cr2O3}. A magnetic proximity effect is pronounced in the first three Pt layers.}
    \label{fig:MPE_number_of_layers_Cr12flipped}
\end{figure}

In the main text a heterostructure with six Pt layers was analyzed. Here, we vary the number of Pt layers to see whether the MPE is affected, in particular to determine whether the thickness of the penetration depth of the MPE depends on the thickness of Pt.
Fig. \ref{fig:MPE_number_of_layers_AFM} and Fig. \ref{fig:MPE_number_of_layers_Cr12flipped} show the magnetization in Pt for up to eight Pt layers for a heterostructure with antiferromagnetic \ce{Cr2O3} and a heterostructure with \ce{Cr2O3} with flipped Cr12 magnetic moment, respectively. Regardless of the magnetic order in \ce{Cr2O3} and regardless of the thickness of Pt, the MPE is pronounced only in the first three Pt layers adjacent to \ce{Cr2O3}. As described in Sec. \ref{sec:magnetic}, Pt is coupled antiferromagnetically to the Cr12 interface magnetic moment and the flipped Cr12 magnetic moment increases the MPE, which explains the change in sign as well as the difference in magnitude between Fig. \ref{fig:MPE_number_of_layers_AFM} and Fig. \ref{fig:MPE_number_of_layers_Cr12flipped}.

\FloatBarrier
\bibliography{references}

\end{document}